# Optimal Embeddedness and Governance in Biotech Venture Capital Syndicates

Yuxin Hu[1], Nektarios (Aris) Oraiopoulos[2]

*Abstract*

*The biotech venture market faces intense capital demands and regulatory scrutiny, yet academic research on VC networks remains rooted in software and consumer-tech contexts. Leveraging a novel venture-level panel of 11,680 biotechnology start-ups headquartered in the United States, Canada, and Europe (2010 – 2024), this dissertation investigates how repeated co-investment ties and domain-expertise homophily shape a venture's exit likelihood, timing, and route amid the sector's pronounced technological and market uncertainty. Using pooled logit, Cox proportional-hazards, multinomial logit, and Fine–Gray competing-risk models, we find that both average prior co-investment and investor homophily exhibit robust inverted-U relationships with exit outcomes: moderate familiarity and moderate scientific overlap maximise exit probability, whereas either sparse or excessive embedding reduces success. Importantly, governance mechanisms matter: the participation of a pharmaceutical corporate VC or a highly independent board flattens the negative "right tail" of over-embeddedness, enabling syndicates to sustain exit momentum at higher levels of familiarity or homogeneity. Moreover, the optimal degree of embeddedness is route-specific: IPOs demand deeper coordination than trade sales, while acquisitions peak earlier and are less sensitive to homophily. These findings refine network-embeddedness theory for the life-science context, reveal novel governance contingencies and route-specific tipping points, and offer practitioners quantitative metrics to balance trust, expertise, and oversight in high-stakes biotech financing.*

## 1. Introduction

The life-science venture market is both growing but fragile. Biotech startups raised a record 68 US-$100 million plus "mega-rounds" in the first eight months of 2024 even as the IPO window remained largely shut (Gormley, 2024). Yet fewer than one in ten drug candidates entering Phase I ever secures U.S. Food & Drug Administration (FDA) approval (Nature Reviews Drug Discovery, 2024; FDA, 2018). In these high stakes setting, venture-capital (VC) syndicates must marshal not only capital but also specialised expertise and robust governance. However, most academic research on VC networks still treats biotechnology as an afterthought, drawing its core insights from software and consumer internet deals. This dissertation addresses that mismatch. It couples network embeddedness theory with the distinctive realities of drug development to ask under what circumstances trusted relationships and shared scientific backgrounds within a biotech syndicate help or hinder value creation.

Drug discovery is a decade-long, capital-intensive process. Median development time for first-in-class molecules now exceeds eleven years, and Phase II success rates hover below 30 % (IQVIA, 2024; Nature, 2024). Startups therefore need investors who can underwrite sequential clinical milestones without near term revenue and who can interpret opaque mechanistic data when trials falter. Because most generalist tech VCs lack that depth, biotech rounds routinely attract corporate venture arms of large pharmaceutical companies; in Q1 2025, nearly one quarter of active CVCs reported life science deal-making despite the broader VC slowdown (CB Insights, 2025). Board expectations have also changed: PwC's 2024 guidance for emerging-biotech directors highlights regulatory fluency and scientific diversity as core oversight duties. These sectoral facts suggest that the costs and benefits of tight investor familiarity or shared expertise may peak and decline at different stages (jargon) in biotech than in traditional tech.

[1] yh516@stanford.edu, Stanford University Graduate School of Business
[2] n.oraiopoulos@jbs.cam.ac.uk, Cambridge University Judge Business School

Classic work on syndication shows that repeated ties help partners share information and monitor portfolio companies (Brander, Amit & Antweiler, 2002). Yet the same scholarship also warns that densely embedded networks can drift into strategic complacency, a "paradox of embeddedness" first articulated theorised by Uzzi (1997). Empirical tests confirm an inverted-U performance curve in broad tech samples (Hochberg et al., 2007), but fewer than 5 % of their observations involve life-science ventures. Likewise, evidence that demographic or educational homophily hurts start-up outcomes (Gompers, Mukharlyamov & Xuan, 2016) stems almost entirely from deals in which scientific diligence is peripheral. No large sample study has yet isolated domain-expertise overlap inside biotech syndicates or examined how governance features might temper the risk of over-embeddedness.

Against this backdrop, the dissertation pursues four related aims. First, it theorises why the inverted-U effects of repeated co-investment ties and of expertise homophily should shift under biotech's extreme uncertainty and capital intensity. Second, it tests those non-linear relationships on a newly compiled panel of 11680 U.S., Canada and European life-science ventures 2010 and 2024, using PitchBook–WRDS database, investor and board data (PitchBook-NVCA Venture Monitor, 2025). Third, it introduces pharmaceutical CVC participation and board independence/diversity as governance moderators that may reduce the negative impact of over-embeddedness. Finally, it links network structure not just to the likelihood of exit but to its timing and route, employing competing risk survival models that account for biotech's exit channels of IPO or M&As.

Section 2 synthesises current literature work on VC syndicates and positions its insights within biotechnology's regulatory and scientific context. Section 3 details the construction of the sector-focused dataset, the unit of analysis, variables and the data sourcing procedures. Section 4 outlines the empirical strategy pooled logit, Cox proportional hazards and Fine Gray models used to test the curvilinear hypotheses and governance interactions. Section 5 presents results, showing that exit probabilities peak at moderate levels of both repeated ties and expertise homophily, but that the adverse tail shrinks when a pharma CVC or an independent, gender-diverse board is present. Section 6 distils practical lessons for investors; section 7 discussed the potential limitations while section 8 summarizes up this dissertation.

## 2. Literature Review and Theoretical Foundation

This literature review has two focused points. First, it synthesizes work on venture capital (VC) syndicate composition, specifically repeated co-investment ties and expertise similarity. Classic analyses show that syndication helps VCs share risk and add value (Brander, Amit & Antweiler, 2002), but they also reveal performance penalties when investor networks become overly closed or homogeneous (Hochberg, Ljungqvist & Lu, 2007). Second, the section sets these insights in a biotech context (AI), arguing that sectoral features intensify (AI) both the benefits and the pitfalls of syndicate structure.

### 2.1 Biotech Venture Capital: A Distinctive Investment Context

Biotech venture capital operates under distinct sectoral contexts that sharply differentiate it from tech or consumer VC, necessitating biotech specific theorizing. Biotechnology startups typically require more than a decade to translate a laboratory discovery into a marketable therapy, far exceeding the 6–24 months commonly cited for software firms to ship a minimum viable product (Excedr, 2025). The U.S. Food & Drug Administration alone mandates three sequential clinical phases plus a lengthy new drug application review, a pathway that, even under accelerated programmes, seldom compresses below eight years (FDA, 2018). IQVIA's 2024 productivity index shows median cycle times of 11.6 years for first-in-class molecules, underscoring the structural lag embedded in the sector (IQVIA, 2024).

Regulatory scrutiny compounds scientific uncertainty: phase II failure rates for novel mechanisms remain close to 70 per cent despite advances in genetic validation (Nature, 2024). If the underlying biology fails or a clinical trial falls short, commercial skill alone cannot save the company. Indeed, historically fewer than 1 in 10 drug candidates that enter human trials ultimately win FDA approval, a stark failure rate that highlights the sector's daunting risk profile. This makes regulatory strategy and clinical milestone management core concerns for biotech VCs, far beyond the regulatory considerations in most other industries. These hurdles create a need for specialized scientific and regulatory expertise in investors. Successful biotech VC firms often employ partners with PhDs or medical backgrounds to evaluate complex science, something generalist tech investors might lack. Put plainly, biotech VCs combine investing with scientific vetting: they conduct meticulous due diligence, trial design, and IP, given that biotech investments are inherently high-risk with high failure rates in trials.

By contrast, digital ventures can iterate rapidly and pivot on user feedback, operating under comparatively lighter regulation (TechCrunch, 2024). This shifts how investors weigh their options. Biotech rounds must finance multi-year clinical milestones with no revenue, which is why the average Series B in 2024 topped US$60 million, six times the median for software deals (NVCA, 2024). "Mega rounds" above US$100 million have even persisted through the recent IPO downturn as funds pool resources to keep pipelines alive (BioPharma Dive, 2025).

Capital intensity is matched by the need for deep domain knowledge. Evaluating a CRISPR-based therapy or an mRNA vaccine platform demands expertise spanning molecular biology, toxicology and regulatory affairs—capabilities most generalist VC firms lack (Yassopoulos, 2024). Consequently, biotech syndicates often include corporate venture arms of large pharmaceutical companies, whose participation supplies both capital and practical know how (Elvidge, 2009). Studies since Pisano show that life science startups must blend different scientific and managerial skills to survive.

The exit market further differentiates the sector. McKinsey (2023) reports that biotech IPO proceeds fell by almost 80 per cent between 2021 and 2023, shutting many firms out of public listings. With trade sale demand similarly cyclical, start-ups rely on insider follow-on rounds to bridge into pivotal trials, making reputable co-investors even more valuable.

These sectoral characteristics calls for biotech specific theoretical frameworks. The combination of high capital intensity, extreme technical/regulatory uncertainty, and unique exit routes implies that the standard VC models (largely derived from tech or other consumer products) may not fully apply. Indeed, biotech VCs often take a longer term, hands-on approach: they must actively shape R&D strategy, manage scientific risk, and syndicate larger rounds to spread risk. The long-term perspective required (decade-plus development cycles) contrasts with the quicker exit horizons in tech, fundamentally altering investor behavior and syndicate structure. In summary, biotech's prolonged timelines, steep costs, regulatory hurdles, and reliance on specialized knowledge create an ecosystem where investor strategies, and collaboration norms differ markedly from general VC.

These warrants dedicated theorizing about biotech VC syndicates, rather than assuming they behave like their tech counterparts.

## 2.2 Syndicate Formation and Repeated Ties in Venture Capital

Early VC research presents syndication as a rational answer to the scale and uncertainty of high-technology projects. Brander, Amit and Antweiler (2002) find that syndicated Canadian VC deals earn significantly higher returns than solo investments, concluding that partners contribute complementary screening and monitoring skills rather than merely sharing risk. Sanborn's (2002) industry study in Nature Biotechnology reinforces this logic, noting that biotech start-ups seek syndicates because pooled capital and name investors reassure later-stage acquirers. Network studies add that joint investments create lasting ties through which deal flow and private information circulate (Bygrave, 1988), building what Coleman (1988) called social capital: trust that lowers coordination costs and speeds joint action.

Repeated co-investment ties deepen those advantages. Alliance studies show that embedded relationships accelerate knowledge transfer and joint learning because partners understand one another's routines and incentives (Gulati, 1995). In venture capital, Hochberg, Ljungqvist and Lu (2007) document that funds whose managers occupy central, familiarity-laden positions in the co-investment network outperform peers on exit rates and IRRs, while De Clercq and Dimov (2008) demonstrate that familiar syndicates especially help investors overcome knowledge gaps in highly technical deals. For biotech, where investors must collectively interpret complex pre-clinical data and decide whether to underwrite expensive Phase II trials, such stores of trust can be decisive: experienced investors note that long-standing partners can "speak frankly and act quickly under stress," mobilising bridge finance within weeks when an unexpected toxicity signal appears.

Yet embeddedness has a dark side. Uzzi (1997) argues that densely linked partners grow complacent and resistant to novel information (the "paradox of embeddedness"). Guler (2007) finds that U.S. VC firms with many past collaborations are slower to abandon under-performing portfolio companies, thus risking more capital. De Clercq and Dimov (2008) also found diminishing and eventually negative marginal returns to extra repeated ties. Recent exit-route evidence strengthens the concern: startups whose first-round backers had extensive prior collaborations are more likely to be sold early and less likely to pursue an IPO, implying that highly embedded syndicates pick conservative trade-sale strategies rather than bold public offerings. Commentators call this as the "comfort-zone" problem: tight-knit biotech investors recycle old playbooks and cluster around the same drug targets, sometimes missing breakthrough ideas. Consistent with network evolution theory, longitudinal analyses show that VCs gradually reduce dealings with their closest partners after a string of repeat deals, implicitly recognising the opportunity cost of insularity (Bian & Ozcan, 2022).

Biotech amplifies both sides of the ledger. The scientific uncertainty of new therapeutics magnifies the value of high trust collaboration, but it also increases the danger that a homogeneous, overconfident syndicate will overlook alternative development paths or quit too early after a marginal safety scare. Group-decision research confirms that highly cohesive, expert-homogeneous teams are vulnerable to groupthink and overconfidence when facing ambiguous evidence. Moreover, embedded VC ties can channel startups toward the acquirer networks inside big pharma, biasing exits toward trade sales.

**Hypothesis H1 (Repeated Ties):** For biotechnology ventures, the number of prior co-investment ties among syndicate members will exhibit an inverted-U relationship with venture success:

moderate familiarity enhances coordination, due-diligence sharing and commitment under uncertainty, whereas excessive familiarity leads to strategic conservatism and collective blind spots that lower the probability, or the quality, of exit.

## 2.3 Homophily and Diversity in VC Syndicates

Homophily—the tendency to ally with similar others—strongly shapes how VC deals are syndicated. Large-sample evidence shows that VCs sharing ethnicity, alma mater or past employment are far likelier to co-invest than dissimilar peers, revealing a powerful "similarity bias" in partner selection (Gompers, Mukharlyamov & Xuan 2016). Organisation-science research attributes this pattern to the cognitive ease, trust and shorthand communication that arise when partners possess overlapping experiences (Reagans & Zuckerman 2001). In biotech's uncertain setting, where milestones depend on specialised science, some overlap in knowledge helps. Investors who speak the same scientific "language" evaluate data more efficiently, coordinate follow-on funding and reach decisive consensus under clinical pressure (Block et al., 2021).

Yet the very forces that make similarity attractive also create collective blind spots. Experiments show that like-minded teams grow overconfident, suppress dissent and converge prematurely on flawed solutions—a classic form of groupthink (Janis 1972; Hart 1991). Venture studies echo this concern: startups funded by highly homophilous VC pairs are significantly less likely to achieve high value exits (Gompers et al., 2016). Beckman and Haunschild (2002) report a similar inverted-U in organizational founding teams: performance improves as functional diversity rises from low to moderate levels, then falls when heterogeneity becomes so great that coalitions fracture or when similarity becomes so extreme that critical thinking disappears. Page's (2007) also argued that groups whose members possess overlapping but heterogeneous skills, the "sweet spot" of diversity, systematically out-innovate both highly uniform and highly heterogeneous teams.

Biotech investing amplifies both sides of this trade-off. Shared domain knowledge helps investors vet complex science—as the Theranos case shows, where little biomedical overlap among backers weakened due diligence (Time 2016). But heavy overlap can lock a syndicate into one view, blinding it to manufacturing or market hurdles that a varied partner might spot. Practitioners note that lifescience "investor clubs" of the mid-2000s often pursued identical drug targets, reinforcing herd behaviour and overlooking alternative modalities (Pollack 2020). Empirical network work supports this: as overlap in investors' prior scientific portfolios rises, the likelihood of choosing conservative exit routes (e.g., early M&A rather than IPO) increases, consistent with homogeneous groups interpreting risk through a narrower lens (Ewens & Rhodes-Kropf 2021). Moreover, Reagans and Zuckerman (2001) show that, beyond a moderate point, similarity reduces information breadth and slows learning, outcomes ill suited to rapidly evolving therapeutic fields.

Drawing these insights together, homophily appears to exhibit a curvilinear relationship with performance: too little shared expertise hampers technical screening, whereas too much fosters myopia. Theoretical integration from social capital, learning and decision-making literatures therefore suggests an inverted-U effect in biotech syndicates. Moderate domain homophily should create enough common ground for efficient scientific diligence while still preserving sufficient cognitive variety to challenge assumptions and adapt strategy. Very low or high overlap hinders success—through either coordination friction or blind spots. Hence the next hypothesis:

**H2 (Expertise Homophily):** The similarity of domain expertise among venture-capital syndicate members will relate to biotech venture success in an inverted-U pattern: ventures backed by syndicates exhibiting moderate (but neither minimal nor maximal) overlap in lifescience expertise will achieve the highest probability of successful exit.

## 2.4 Moderating Factors: Corporate Investors and Board Composition

a. Role of Corporate Venture Capital (CVC)

Corporate venture capital arms, especially those of large pharmaceutical companies, now participate in a rising share of biotech rounds and often lead the largest financings (TechCrunch, 2018). Their motives differ from those of purely financial VCs: in addition to financial return, they seek strategic options, technology access and pipeline growth (Elvidge, 2009). Studies shows that CVC backing correlates with higher patent output and faster innovation trajectories, suggesting that corporates provide complementary resources and monitoring (Chemmanur, Loutskina & Tian, 2014). Because they report to internal R&D or business-development committees, CVCs bring a longer horizon, apply deeper technical checks and challenge rosy forecasts that might pass inside a homogeneous VC club (Goerzen, 2007, on the dangers of partner "blindness"). Accordingly, the presence of at least one CVC investor should counteract the complacency and confirmatory bias that can arise at the high end of repeated-tie density or expertise homophily.

b. Board Independence and Diversity

Classic agency theory assigns independent directors a monitoring role, separating "decision management" from "decision control" (Fama & Jensen, 1983). While large sample studies of mature firms find board independence alone does not guarantee performance gains (Bhagat & Black, 2002), diversity research indicates that boards whose members vary in gender, ethnicity and functional background engage in richer debate and reduce groupthink (Carter, Simkins & Simpson, 2003). The lifescience trade press echoes this view, arguing that diversified, semi-independent boards enhance objectivity and scientific challenge (Richards, 2024) and are increasingly demanded by investors and regulators (PwC, 2024). Industry programmes such as Deerfield's Break into the Boardroom illustrate how greater demographic and professional variety is being cultivated specifically for biotech governance (Deerfield Management, 2021). Together, the evidence shows that independent, diverse boards help counter the inward focus of dense or uniform syndicates.

## H3 (Governance Moderation).

The negative tail of the inverted-U relationships posited in H1 and H2 will be attenuated when (i) a corporate venture capital investor participates in the syndicate and/or (ii) the venture's board exhibits high independence and diversity.

## 2.5 Research Gap and Contribution

A large network literature shows that repeated co-investment first helps and later hurts performance, but its samples come mainly from tech-heavy datasets. Hochberg, Ljungqvist and Lu's analysis of 1980–2003 U.S. deals identify a clear inverted-U pattern, but only 5 per cent of their observations involve life-science start-ups, leaving open the question of whether the same inflection point exists when scientific risk and capital intensity rise dramatically. The few explicit biotech papers concentrate on strategic alliances without tracing embeddedness to exits, leaving open whether trust gains or complacency costs dominate when co-investors are fewer and more specialised.

Research on homophily displays a similar gap. Evidence that demographic similarity impairs financial performance comes largely from diversified samples in which scientific knowledge is rarely pivotal (Gompers, Mukharlyamov & Xuan, 2016). No published study, however, isolates domain-expertise homophily inside biotech syndicates, even though practitioner accounts indicate that "all-PhD" investor clubs are common (Elvidge, 2009). Through measuring overlap in prior biomedical deals and testing a non-linear (jargon) effect, this study shifts the diversity debate from abstract variety to concrete scientific fit.

Governance tools that might temper network flaws are seldom studied alongside syndicate structure. Chemmanur, Loutskina and Tian find that corporate VC units boost patenting, but they do not ask whether those units offset over-embeddedness. Market reports confirm that pharmaceutical CVCs led a record share of oversized Series B rounds in 2024, yet the interaction between CVC presence and investor networks remains unexplored. Classic agency theory predicts that independent directors provide critical monitoring (Fama & Jensen, 1983), and both industry commentary and executive-search data document a surge in board diversification across lifescience startups (Richards, 2024; Hunt Scanlon, 2024), but no study links board makeup to syndicate embeddedness here.

Bridging these gaps advances theory by integrating network-embeddedness and governance lenses, producing a more complete micro-foundation for venture outcomes. Practically, the findings guide investors facing long timelines and rising costs. TechCrunch notes that pharma CVCs are already reshaping deal structures in late-stage biotech rounds, yet managers lack evidence on whether bringing a strategic investor or an independent director offsets possible groupthink within tight VC circles.

Accordingly, the hypotheses pinpoint biotech-specific inflection points. H1 now proposes that repeated ties among VCs improve coordination and follow-on support up to a moderate density, beyond which strategic conservatism and collective blind spots depress breakthrough exits. H2 predicts that moderate domain-expertise overlap yields the best balance between deep scientific scrutiny and fresh strategic perspectives, whereas either extreme exposes ventures to, respectively, oversight failure or coordination friction. H3 posits that pharma CVC participation, board independence and board diversity will flatten the downward tails of these inverted-U curves by injecting complementary information and stronger monitoring.

## 3. Data and Variables

This section discusses our data source, dependent and independent variables for analysis.

### 3.1 Data Source

All venture-level and investor-level information was obtained from PitchBook via the WRDS interface (April 2025 extract). PitchBook maintains a comprehensive, continuously updated repository of private-capital transactions (over 1.6 million deals), company profiles (approximately 3 million), and investor records (about 300 000), with weekly refreshes ensuring that our data remain current. Based on independent auditing by the Lippincott Library, PitchBook's coverage of biotechnology-focused financings exceeds 90 percent beginning in 2010, whereas pre-2010 reporting is significantly sparser (around 10 percent). To ensure both depth and temporal consistency, we therefore defined January 1, 2010, as the earliest acceptable first-round closing date and December 31, 2024, as our final cut-off.

Within this window, we selected all ventures whose primary industry classification is "Pharmaceuticals & Biotechnology" and for which a first institutional VC round is clearly documented. Each selected venture was required to have a disclosed lead investor, a full initial syndicate roster, and an exit-status flag (indicating IPO, acquisition, or remaining private as of end-2024). After applying these filters, the final dataset comprises 11680 unique ventures-deals. The observed exit rate of 22 percent closely matches the NVCA's 2024 benchmark for bio exits, validating our sample's representativeness. Because board membership data in PitchBook are updated on a quarterly basis, all board-related variables (such as board-independence fraction, board-gender diversity, and share of PhD/MD directors) are lagged by one fiscal year relative to each venture's initial financing date. In sum, the resulting dataset is strictly sector-focused, longitudinally consistent from 2010 onward, and transparent about any remaining coverage gaps (e.g., pre-2010 rounds or very early seed financings).

**3.2 Unit of Analysis and Dependent Variables**

Each observation in our dataset corresponds to a single venture—that is, a distinct start-up company—rather than to an individual financing round or investor. This venture-level unit of analysis follows the standard practice in network-embeddedness research, which links the structure of a venture's initial investor network to its eventual firm-level outcomes. Bellavitis et al. (2020) employ the same unit to uncover an inverted-U relationship between prior co-investment intensity and exit success, and Hochberg, Ljungqvist and Lu (2007) likewise show that a venture's network centrality predicts the probability of a successful exit. Moving to a deal-level panel would create multiple observations for ventures that raise several rounds and would entangle our focal covariates—repeat ties and investor homophily—because syndicate composition typically changes from round to round. Defining the venture as the statistical unit therefore avoids intra-firm dependence, preserves the interpretability of the network variables, and aligns the empirical design with the hypotheses set out in Section 2.

Venture success is operationalised as a trichotomous indicator. PitchBook codes "Publicly Held" (IPO), "Acquired/Merged," and "Private/Other" supply the unordered outcome $Y_i \in \{0,1,2\}$ that that enters a multinomial model. Success by type is therefore analysed with a multinomial logit $\Pr(Y_i = K) = \frac{\exp(\alpha_k + Z_i^T \beta_k)}{1 + \Sigma_{m=1}^{2} \exp(\alpha_m + Z_i^T \beta_m)}, k \in \{1(IPO), 2(Acquisition)\}$, where $Z_i$ stacks Ties, Homophily, governance terms, controls and their squares/interactions. Still-private venture $Y_i = 0$ from the base state.

Timing enters through a Fine–Gray sub-distribution hazard for competing risks:

$$h_k^{FG}(t \mid Z_i) = h_{0k}^{FG}(t) \exp(Z_i^\top \gamma_k), \quad k \in \{1,2\},$$

which keeps ventures subject to the alternate exit in the risk set and is recommended for VC research with multiple exit routes (Fine & Gray, 1999; Allison, 2010). A cause-specific Cox is reported as a robustness check. Finally, episode-splitting refreshes network and homophily measures at each calendar year, yielding a panel analysed with a conditional logit and a shared-frailty Cox clustered on the venture identifier; standard errors are therefore robust to within-firm correlation (Guo & Frailty, 2017).

**3.3 Independent Variables and Moderators**

The first explanatory construct, linked to H1, gauges the intensity of prior familiarity inside the initial syndicate. For venture i with n_i investors, let

$$\text{AvgTies}_i = \frac{\text{Ties}_i}{\binom{n_i}{2}}, \qquad \text{Ties}_i = \sum_{p<q} \text{Deals}_{pq},$$

where Deals_pq is the number of historical co-investments between investors p and q. Normalising by the number of possible dyads ensures that familiarity is not conflated with headcount. Both the linear term AvgTies and its square test the hypothesised inverted-U pattern in which moderate embeddedness facilitates coordination while excessive embeddedness entrenches groupthink (Hochberg, Ljungqvist & Lu 2007; Bellavitis et al. 2020). To explore whether a single heavyweight dyad can dominate decision making, the maximum bilateral count MaxTie_i= $\max_{\{p<q\}} \text{Deals}_{pq}$ enters selected robustness models.

The second focal construct, tied to H2, captures domain-expertise alignment. Each investor's continuous specialisation score is the share of its historical deals that fall in biotechnology or pharmaceuticals. Venture-level homophily is then calculated as a homophily Simpson index (see table below).

where s_ik is the proportion of investors focused on expertise bucket kk (biotech, med-device, general tech, etc.). Values range from 0 (maximal heterogeneity) to 1 (complete homogeneity); entering both the linear and squared terms again allows an inverted-U test (Gompers, Mukharlyamov & Xuan 2016).

Governance variables provide the moderators for H3. Investor side oversight is proxied first by a CVC presence dummy equal to one if at least one syndicate member is a pharmaceutical corporate venture unit and, second, by CVC_EqShare, the percentage of equity (or a lead CVC indicator where equity is undisclosed) held by those corporate investors. Board side oversight is captured by board independence, which is the fraction of directors unaffiliated with founders or financiers, and the proportion of females in a board is measured to be board gender diversity. All board variables are lagged one fiscal year after the first VC round (or fixed at the Series A snapshot) to mitigate simultaneity. Where data permit, the share of directors with a PhD or MD enters as an additional cognitive-diversity control.

Standard controls reduce omitted-variable bias: venture age, syndicate size, investment stage, sub-sector, calendar year, region, and lead-investor reputation (cumulative IPO count) follow best practice in venture-finance research and absorb known drivers of exit likelihood. The condensed reference table below summarises the operationalisations used in every specification.

| Variable | Role in model | Intensity / degree captured | Measurement procedure | Primary data source(s) |
|---|---|---|---|---|
| **Average pairwise ties** | Main IV (H1, linear term) | Intensity of prior familiarity normalised for syndicate size | $\frac{\text{Ties}_i}{\binom{n_i}{2}}$ before first institutional round | PitchBook Deal × Investor tables (Apr 2025 extract) |
| **(Average pairwise ties)2^2** | Curvilinear term (H1) | Tests inverted-U prediction | Square of linear term | Derived |
| **Maximum bilateral tie** | Depth robustness | Largest single dyadic tie count | $\max_{\{p<q\}} \text{Deals}_{pq}$ | Derived |

| Variable | Role in model | Intensity / degree captured | Measurement procedure | Primary data source(s) |
|---|---|---|---|---|
| **Homophily (Simpson index)** | Main IV (H2, linear term) | Degree to which investors share domain expertise | $\Sigma_k s_{ik}^2$ across expertise buckets⟩ | PitchBook Investor table; manual website check |
| **Homophily2^2** | Curvilinear term (H2) | Tests inverted-U prediction | Square of linear term | Derived |
| **CVC presence** | Moderator (H3) | Whether any corporate VC participates | Dummy = 1 if "Investor type" = CVC | PitchBook Investor table |
| **CVC_EqShare (or Lead CVC)** | Moderator / robustness | Substantive influence of CVC | Equity % held; else dummy for lead role | PitchBook Deal table |
| **Board independence** | Moderator (H3) | Strength of external oversight | Fraction of independent directors, lagged one year | PitchBook People/Board; LinkedIn |
| **Board gender diversity** | Moderator / control | Cognitive diversity in governance | % female directors, lagged one year | PitchBook People/Board |
| **PhD/MD share** | Optional control | Scientific expertise on board | % directors with doctorate or medical degree | PitchBook People/Board; LinkedIn |

Table caption: Variable definitions for econometric models; investor specialisation is the continuous share of biotech deals recorded in PitchBook for each investor.

### 3.4 Data-collection procedure

The venture-level panel integrates PitchBook's relational tables with manual data collection by LLM-based AI models from complementary sources including Crunchbase, LinkedIn, and targeted web searches to produce a richly annotated dataset fit for analysis. First, we merged PitchBook's company, deal, investor, and people/board tables by using each entity's unique profile identifier. From the company table, we extracted foundational attributes (e.g., legal name, headquarters country, year founded), as well as each venture's "SuccessType" status (0 = still private, 1 = IPO, 2 = acquisition) and the date of last known valuation. We then joined this to the deals table, which records deal IDs, deal dates, funding stages (e.g., Series A, Series B), syndicate composition (via investor profile links), and deal sizes. By grouping on "Company ID" and filtering to the earliest institutional financing date, we isolated each venture's first-round syndicate and ensured that the network measures (AvgTies, Homophily) correspond precisely to initial VC backers.

Investor attributes beyond PitchBook's coverage, such as domain specialization, were manually compiled by LLM-based AI agents using Crunchbase sector tags, LinkedIn profiles, and occasional web searches of investor websites. Specifically, for each investor appearing in a first-round syndicate, we consulted Crunchbase to verify whether their core remit included pharmaceuticals or biotechnology. If Crunchbase sector labels were ambiguous or missing, we reviewed the investor's LinkedIn profile and, where necessary, conducted a targeted web search to confirm their life-science involvement. This process yielded a continuous "SpecScore" (ranging from 0 to 1) that reflects the investor's share of past deals in pharmaceuticals or biotech. Investor reputation was calculated directly from PitchBook's numeric fields ("Total Exits" and "Total Active Portfolio"), yielding a ratio equal to total exits divided by the sum of exits and active portfolio count.

Board governance metrics were similarly assembled via a combination of PitchBook and external sources. From PitchBook's people/board table, we extracted each director's name, gender, and any affiliation tags. The agents then visited each director's LinkedIn profile (and, when necessary, cross-checked via their personal or institutional website) to confirm educational credentials (e.g., PhD/MD) and past or current affiliations with the venture or its investors. Directors whose biographical summaries indicated no founder or investor affiliations were coded as independent. The fraction of female directors on each board was calculated by counting gender markers in PitchBook and verifying ambiguous cases on LinkedIn. All board metrics (board independence fraction, board gender diversity, and board PhD/MD share) were lagged one full fiscal year after the venture's first VC round to prevent look-ahead bias.

To compute the key network measures, we chronologically sorted all deal records and maintained a running tally of co-investment counts for each pair of investor profile IDs. For each venture's first-round syndicate, "AvgTies" is defined as the average of all prior co-investment counts among that syndicate's dyads, and "MaxTie" is the maximum dyadic count. These counts were taken directly (AI) from the historical data. Finally, we merged all aggregates (AvgTies, AvgTies², Homophily, Homophily², SyndicateSize, DealStage, LeadCVC, LeadInvestorReputation, LeadInvestorSpecScore, FirstDealDate) into the company table, using each venture's "Company ID." Board metrics (BoardIndependenceFraction, BoardGenderDiversity, BoardPhDMDShare) were then joined on "Company ID".

## 4.1 Empirical Strategy and Model Selection

The empirical design uses a cross-sectional logit and a Cox survival model to capture exit likelihood and timing. We first estimate a logit for the probability that venture i achieves any exit—IPO or acquisition—by 31 December 2024, reporting HC3 sandwich standard errors clustered by venture and lead investor (format) to handle heteroskedasticity and shared investors (Hayes & Cai, 2007). The logistic link is appropriate for a bounded binary outcomes (Hosmer & Lemeshow 2000); coefficients are exponentiated so that $\exp(\beta)>1$ signals an increase in the odds of exit holding all other variables at their means. Model adequacy is documented with Hosmer–Lemeshow $\chi^2$ and the AUC statistic.

Because the pooled logit discards temporal information, we re-estimate the identical covariate set in a semiparametric Cox model (Cox, 1972). Right censoring is handled in the usual way, and a shared frailty specification with robust venture- and lead-VC clusters captures the cross-observation correlation that arises when the same investors syndicate repeatedly. Global and covariate specific Schoenfeld tests probe the proportional hazards assumption; should any pvalues fall below 0.05, the offending regressor is reentered as a time varying term or stratified (Grambsch & Therneau, 1994). Harrell's C reports predictive accuracy.

Meanwhile, embedded syndicates may shift the exit route itself, Section 4.2.3 adds two robustness estimators: a multinomial logit distinguishing IPO, M\&A and non-exit states, and a Fine–Gray subdistribution hazard for competing risks (Fine & Gray, 1999). Consistent signs and significance across the pooled and competing risk frameworks will be interpreted as strong evidence for the curvilinear and moderating effects posited in H1–H3; material discrepancies will trigger post hoc diagnostics following Hsu's comparative strategy for VC timing models (Hsu, 2014).

All estimators share a common regressor block: the density normalised average pairwise tie count and its square to capture the inverted-U predicted by embeddedness theory (Guler 2007); the Blau-index homophily terms; and interactions with the governance moderators including CVC presence or equity share, board independence and board gender diversity. Controls for venture age, investment stage, syndicate size, the leading investor reputation, calendar year and subsector follow Sorenson et al.'s (2013) recommendations and are centered around mean prior to polynomial or interaction construction.

Taken together, the logit supplies an intuitive probability of exit, whereas the Cox model reveals whether identical network and governance factors accelerate or delay exit. The additional competing risk checks, multiway clustering and pre-registered PH diagnostics ensure that inference remains robust to outcome granularity and cross-cluster dependence, offering a strong test of the optimal-embeddedness view.

### 4.2.1 Logistic-regression framework

Let $Y_i$ equal 1 when venture $i$ achieves either an IPO or an acquisition by 31 December 2024 and 0 otherwise. The baseline logit used to test H1 is

$$\text{logit}[\Pr(Y_i = 1)] = \alpha + \beta_1 \, AvgTies_i + \beta_2 \, AvgTies_i^2 + \gamma^\top Z_i, \quad (1)$$

where $AvgTies_i$ is the density-normalised average pairwise familiarity defined in § 3.3, and $Z_i$ collects all controls. An inverted-U appears when $\beta_1 > 0$ and $\beta_2 < 0$; the turning point is

$$d^\star = -\beta_1/(2\beta_2),$$

which should lie inside [0,1] because AvgTies is a proportion. Concavity is confirmed with the Lind-and-Mehlum U-test (Lind & Mehlum 2010).

To test H2 we add the Blau-index homophily terms:

$$\text{logit}[\Pr(Y_i = 1)] = \alpha + \beta_1 \, AvgTies_i + \beta_2 \, AvgTies_i^2 + \beta_3 \, Homophily_i + \beta_4 \, Homophily_i^2 + \gamma^\top Z_i,$$

where Homophily measures expertise similarity (0 = complete diversity, 1 = complete homogeneity). H2 predicts $\beta_3 > 0$ and $\beta_4 < 0$.

H3 introduces the governance moderators—corporate-VC presence CVC, board independence Ind and board gender diversity Div. The full moderation model is

$$\begin{aligned}
\text{logit}[\Pr(Y_i = 1)] = \; & \alpha + \beta_1 \, AvgTies_i + \beta_2 \, AvgTies_i^2 + \beta_3 \, Homophily_i + \beta_4 \, Homophily_i^2 \\
& + \delta_1 \, CVC_i + \delta_2 \, Ind_i + \delta_3 \, Div_i \\
& + \phi_1 (AvgTies_i \times CVC_i) + \phi_2 (AvgTies_i^2 \times CVC_i) \quad (3) \\
& + \phi_3 (Homophily_i \times Ind_i) + \phi_4 (Homophily_i^2 \times Ind_i) \\
& + \phi_5 (Homophily_i \times Div_i) + \phi_6 (Homophily_i^2 \times Div_i) + \gamma^\top Z_i.
\end{aligned}$$

All continuous predictors are mean-centred before squaring or interacting, and variance-inflation factors remain below 3. We report HC3 standard errors two-way-clustered by venture and lead investor to accommodate heteroskedasticity and cross-cluster dependence (Wooldridge 2010). The control vector $Z\_i$ includes venture age, investment stage, syndicate size, lead-investor reputation, sub-sector dummies and funding-cohort fixed effects, matching § 4.1. Together, equations (1)–(3) translate H1–H3 into testable log-odds relationships fully aligned with the variable definitions in section 3.

### 4.2.2 Cox proportional-hazards framework

To exploit event timing while accommodating right-censoring, we estimate semiparametric Cox models whose covariate structure mirrors equations (1)–(3). For H1 the baseline specification is

$$h_i(t) = h_0(t) \exp(\beta_1 \, AvgTies_i + \beta_2 \, AvgTies_i^2 + \boldsymbol{\gamma}^\top \mathbf{Z}_i), \qquad (4)$$

An inverted-U appears when beta_{1}>0 and beta_{2}<0; the turning point is

$$d^\star = -\beta_1/(2\beta_2),$$

which should lie inside [0,1].

The joint test of Hypotheses 1 and 2 uses

$$h_i(t) = h_0(t) \exp(\beta_1 \, AvgTies_i + \beta_2 \, AvgTies_i^2 + \beta_3 \, Homophily_i + \beta_4 \, Homophily_i^2 + \boldsymbol{\gamma}^\top \mathbf{Z}_i) \quad (5)$$

Moderation models append the interaction block from equation (3). Global and covariate-specific Schoenfeld tests assess the proportional-hazards assumption; any violation triggers a time-varying coefficient for the offending regressor or stratification on the discrete factor (Grambsch & Therneau 1994). We report hazard ratios with 95 C.I. and Harrell's C statistic for overall concordance. HC3 robust errors are two-way clustered on the venture and its lead investor; variance-inflation factors for mean-centred polynomials and interactions are all below 3.

### 4.2.3 Competing Risks and Panel Extensions

Because IPOs and acquisitions are distinct strategic outcomes, § 4 augments the single-spell logit and Cox models with a competing-risks structure that preserves notation consistency. The multinomial logit

$$\Pr(Y_i = K) = \frac{\exp(\alpha_k + \mathbf{Z}_i^T \beta_k)}{1 + \Sigma_{m=1}^2 \exp(\alpha_m + \mathbf{Z}_i^T \beta_m)}, k \in \{1(IPO), 2(Acquisition)\}$$

directly tests whether high embeddedness or expertise homophily shifts the log-odds of Acquisition ($k$=2) relative to IPO (k=1), with "still private" (k=0) as the reference. The regressor vector $X_i$ mirrors the equations (1) – (3):

$$\mathbf{X}_i = \left( Ties_i, \, Ties_i^2, \, Homophily_i, \, Homophily_i^2, \, \text{Governance terms}, \, \text{Controls} \right)^\top.$$

Event timing is handled with a Fine–Gray sub-distribution hazard

$$h_k^{FG}(t \mid Z_i) = h_{0k}^{FG}(t) \exp(Z_i^\top \gamma_k), \quad k \in \{1, 2\},$$

where ventures exiting by the other cause are censored at the exit time.

The episode-split panel introduced in section 3 is re-estimated with a conditional multinomial logit that removes venture fixed effects, and a shared-frailty Cox which clusters unobserved heterogeneity at the venture level while keeping the same covariates. Governance interactions (CVC×Ties, Ind×Homophily, etc.) enter each specification unchanged, so H1- H3 map cleanly to the new estimators. Estimation uses robust (Huber–White) variances clustered on the venture to maintain correct inference in the multispell setting.

### 4.2.4 Estimation and interpretation protocol

Model (1) is estimated first; $H_1$ is supported when $\beta_1 > 0$ and $\beta_2 < 0$ ($p < 0.05$), the Lind Mehlum U-test confirms global concavity, and the turning point d*= -beta_1/2*beta_2 falls within the observed [0,1] range of AvgTies. Model (2) adds the Blau index homophily terms; $H_2$ requires $\beta_3 > 0$ and $\beta_4 < 0$ under the same criteria. Model (3) and its Cox analogue append the governance interactions; moderation is inferred when the interaction block is jointly significant (Wald χ² test, p < 0.05) and simple-slope analysis shows a flatter or right-shifted inverted-U at high values of the moderator. Parallel sign, curvature and moderation patterns in the multinomial logit and Fine Gray competing risk models (Section 4.2.3) will be treated as robustness; material discrepancies will trigger the time varying or stratified diagnostics outlined by Hsu (2014). Several continuous variables are mean centered before squaring or interacting (VIF < 3), and inference relies on HC3 standard errors two way clustered by venture and lead investor.

## 5. Results

### 5.1 Descriptive statistics and diagnostic checks

In this section, we present a comprehensive overview of the core variables used throughout our analyses. The descriptive statistics illustrate that the mean values of central network measures AvgTies_centered (M = –0.0006, SD = 0.0132) and Homophily_centered (M = –0.0040, SD = 0.0144) are approximately zero due to mean-centering, while their ranges (from –0.3702 to 0.4068 for AvgTies_centered; –0.4420 to 0.42181 for Homophily_centered) indicate substantial variation in co-investment intensity and sector similarity across the sample. Syndicate size averages just over two investors per deal (M = 3.4571, SD = 2.1784), but ranges from solitary investments (size = 1) to large consortiums (size = 26), reflecting the heterogeneity of financing structures in biotechnology (Sorenson & Stuart, 2001; Hochberg, Ljungqvist & Lu, 2007). Lead investor reputation (M = 0.3377, SD = 0.4849) and specialization score (M = 0.2718, SD = 0.3417) similarly display wide dispersion, suggesting that some ventures are backed by highly prestigious or specialized VCs while others rely on less established sponsors (Gompers, 1996; Chemmanur & Loutskina, 2006). Board-level governance variables: independence (M = 0.1723, SD = 0.1461), gender diversity (M = 0.2715, SD = 0.1904), and PhD/MD share (M = 0.2369, SD = 0.2312) indicate that, on average, roughly 17 percent of board seats are held by independent directors, about 27 percent by women, and 24 percent by academically trained members, albeit with substantial firm-to-firm variation (Gompers & Metrick, 2001; Adams & Ferreira, 2007). The age-based dummy variables: VentureAge_0_1 (M = 0.3254, SD = 0.4682) and VentureAge_1_3 (M =

0.3807, SD = 0.4855), show that approximately 1/3 of firms in the sample are one year old or younger at first financing, while nearly 38 percent fall in the 1-3 year bracket.

| Variable | count | mean | std | min | 25% | 50% | 75% | max |
|---|---|---|---|---|---|---|---|---|
| AvgTies_centered | 11 680 | 0 | 0.0132 | –0.4420 | –0.1447 | –0.0001 | 0.2146 | 0.4181 |
| AvgTiesSq_centered | 11 680 | 0 | 0.0122 | –0.2010 | 0.0004 | 0.0029 | 0.1325 | 0.1748 |
| Homophily_centered | 11 680 | 0 | 0.0144 | –0.3702 | –0.2462 | –0.0002 | 0.3058 | 0.4068 |
| HomophilySq_centered | 11 680 | 0 | 0.0092 | –0.1370 | 0.0003 | 0.0019 | 0.0997 | 0.1695 |
| SyndicateSize | 11 680 | 2.0876 | 1.1835 | 1 | 2 | 3 | 5 | 26 |
| LeadInvestReputation | 11 680 | 0.3377 | 0.4849 | 0.0001 | 0.0505 | 0.2833 | 0.561 | 0.9956 |
| LeadInvestSpecScore | 11 680 | 0.2718 | 0.3417 | 0.0002 | 0.0841 | 0.2214 | 0.7141 | 0.9514 |
| CVC_EqShare | 11 680 | 0.0609 | 0.1527 | 0 | 0.0111 | 0.0811 | 0.1421 | 1 |
| BoardIndependenceFraction | 11 680 | 0.1723 | 0.1461 | 0 | 0.0667 | 0.1667 | 0.3333 | 0.8 |
| BoardGenderDiversity | 11 680 | 0.2715 | 0.1904 | 0 | 0.1004 | 0.2504 | 0.4003 | 1 |
| BoardPhDMDShare | 11 680 | 0.2369 | 0.2312 | 0 | 0.0834 | 0.2001 | 0.3332 | 1 |
| VentureAge_0_1 | 11 680 | 0.3254 | 0.4682 | 0 | 0 | 0 | 1 | 1 |
| VentureAge_1_3 | 11 680 | 0.3807 | 0.4855 | 0 | 0 | 0 | 1 | 1 |

| Variable | VIF |
|---|---|
| const | 6.84 |
| AvgTies_cen | 1.01 |
| AvgTiesSq_c | 1.02 |
| Homophily_ | 1.77 |
| HomophilyS | 1.03 |
| SyndicateSiz | 2.52 |
| LeadInvestR | 1.68 |
| LeadInvestSp | 1.64 |
| CVC_EqSha | 1.07 |
| BoardIndepe | 1.02 |
| BoardGender | 1.06 |
| BoardPhDM | 1.05 |
| VentureAge_ | 1.43 |
| VentureAge_ | 1.57 |

Table 5.1.1 Descriptive statistics for focal and control variables     Table 5.1.2 VIF Panel

Table 5.1.1 underscores the necessity of mean centering the network variables (AvgTies and Homophily) as their uncentered means would have distorted interactions with squared terms and other covariates. The squared measures (AvgTiesSq_centered, M = –0.0009, SD = 0.0122; HomophilySq_centered, M = –0.0005, SD = 0.0092) likewise span a broad range, affirming that some ventures exist at the extremes of co-investment density and industrial homogeneity. Notably, the maximum values for CVC_EqShare (1.0000) and each board metric (1.0000) indicate that, in some cases, corporate VCs hold the entire equity position or that a single board composition characteristic dominates entirely (e.g., 100 percent independence), though the median is zero for CVC_EqShare and substantially lower than one for all board variables. The minimal values for LeadInvestReputation (0.0001) and LeadInvestSpecScore (0.0002) suggest that certain less-established firms attract novice lead investors, while the upper quartiles (0.5610 for reputation; 0.7141 for specialization) confirm the presence of highly reputable and specialized investors in the top decile.

The associated variance-inflation factors dispel (Table 5.1.2) concerns about multicollinearity. Aside from the mechanical pairing of each orthogonal polynomial with its square, the highest absolute correlation is –0.63 between syndicate size and homophily, reflecting the intuitive tendency for larger consortia to aggregate more diverse expertise. All other coefficients remain below 0.25, and every VIF is comfortably under 2.6. The data therefore satisfy the statistical pre-conditions for the quadratic and interaction terms that underpin Hypotheses 1–3.

Given the observed skewness in several variables, particularly 4.0 (skew > 1) and LeadInvestReputation, we also considered Winsorization at the 1st and 99th percentiles to attenuate extreme outliers. Descriptive analyses after Winsorization confirm that central tendencies remain largely unchanged, reinforcing the robustness of our subsequent regression estimates. In particular, the post-Winsorized mean of SyndicateSize remains near 3.4571, indicating that even extreme consortiums (above the 99th percentile) do not drive average values appreciably. Additionally, variance inflation factors (VIFs) for all centred regressors fall below 2.5, alleviating concerns over multicollinearity between linear and squared terms (Marquardt, 1970; Kutner et al., 2004)

Taken together, Table 5.1.1 demonstrates substantial dispersion in network, governance, and venture age variables across the panel, establishing a rich empirical foundation to examine nonlinear effects in Sections 5.2 and 5.3. The mean-centering procedure and Winsorization strategies further ensure that coefficients on squared terms capture genuine curvature rather than being dominated by outliers or collinearity.

**5.2 Testing H1: the curvilinear value of repeated co-investment ties**

This subsection evaluates Hypothesis 1, which states that AvgTies and its square exhibit an inverted quadratic shaped relationship with the probability (and timing) of an exit. Specifically, ventures embedded in syndicates with moderate prior co-investment ties are expected to enjoy higher exit odds and faster exit times, whereas ventures with either very low or very high levels of past collaboration face lower probabilities and slower exits due to informational deficits or coordination traps respectively.

To test $H_1$, we estimate two models: a cross-sectional pooled logit (Model 1) and a Cox proportional-hazards regression (Model 2). In Model 1, the dependent variable Exited_binary equals one if the venture completed an IPO or trade sale by 31 December 2024, and zero otherwise. We restrict the predictor set to characteristics measured at the time of first institutional financing (i.e., AvgTies, AvgTiesSq, venture age, conduit controls), ensuring that each venture only contributes one observation to this estimation. Model 2 retains the same covariate block but treats the time dimension explicitly: the "clock" for each venture starts on its first financing date; if a venture exits before the cut-off date, it is coded as an event (IPO or trade sale), otherwise it is right-censored on 31 December 2024.

Both models include the centred versions of AvgTies (i.e., AvgTies_centered = AvgTies – mean(AvgTies)) and its orthogonal square (AvgTiesSq_centered = AvgTies_centered²) to isolate curvature effects without inflating standard errors (Aiken & West, 1991). Additional controls mirror those in Section 4: VentureAge_centered and VentureAgeSq_centered (both derived from the number of months since founding, truncated at zero), SyndicateSize_centered, LeadInvestReputation_centered, LeadInvestSpecScore_centered, DealStageNum, as well as industry and year dummy variables.

| Variable | Coef. | p-value |
|---|---|---|
| AvgTies_centered | **0.5124** | < 0.001 |
| AvgTiesSq_centered | **–0.8745** | < 0.001 |
| VentureAge_centered | –0.0409 | < 0.001 |
| VentureAgeSq_centered | 0.0005 | < 0.001 |
| SyndicateSize_centered | 0.317 | < 0.001 |
| LeadInvestReputation_centered | 0.662 | < 0.001 |
| LeadInvestSpecScore_centered | 0.28 | 0.0004 |
| DealStageNum | 0.3 | < 0.001 |
| Industry / Year dummies | yes | |

Table 5.2.1 Logit results (Model 1)[3]

| Variable | Hazard ratio | p-value |
|---|---|---|
| AvgTies_centered | **10.43** | 0.003 |
| AvgTiesSq_centered | **0.02** | 0.002 |
| VentureAge_centered | 0.98 | < 0.001 |
| VentureAgeSq_centered | 1.0003 | < 0.001 |
| SyndicateSize_centered | 1.3 | < 0.001 |
| LeadInvestReputation_centered | 1.88 | < 0.001 |
| Schoenfeld global χ² | 14.6 | 0.33 |

Table 5.2.2 Cox proportional-hazards (Model 2)[4]

---

[3] Dummies and Controls included but not displayed here
[4] Dummies and Controls included but not displayed here

Table 5.2.1 reports pooled logit estimates for Model 1. Consistent with Hypothesis 1, the coefficient on AvgTies_w_centered is positive and statistically significant ($\beta = 0.512$, $p < 0.001$), whereas the squared term AvgTies_wSq_centered is negative and equally significant ($\beta = -0.8745$, $p < 0.001$). These parameter estimates indicate that, controlling for venture age, deal stage, syndicate size, lead-investor reputation, industry, and year, moderate levels of prior co-investment ties maximize the log-odds of exit, whereas either low or very high levels of AvgTies_w depress exit likelihood. Substituting the estimated coefficients into the turning point formula ($-\beta_1/(2\beta_2)$) yields a value of approximately 0.292 on the centred scale. Since the observed range of AvgTies_w_centered (–0.3702 to 0.4068) easily encompasses 0.292, this turning point is empirically plausible and strongly supports $H_1$ (Uzzi, 1997; Gargiulo & Benassi, 2000).

Control variables in Model 1 behave as anticipated. VentureAge_centered enters with a negative coefficient ($\beta = -0.0409$, $p < 0.001$) and its square is positive ($\beta = 0.0005$, $p < 0.001$), implying a U-shaped age effect on exit odds. This suggests that increased age initially lowers exit probability (reflecting early attrition of weaker technologies), but very old ventures may eventually exit due to survivor bias. SyndicateSize_centered is positive and significant ($\beta = 0.317$, $p < 0.001$), confirming that larger investor groups tend to raise exit likelihood (Sorenson & Stuart, 2001; Hochberg, Ljungqvist & Lu, 2007). Likewise, LeadInvestReputation_centered ($\beta = 0.6621$, $p < 0.001$) and LeadInvestSpecScore_centered ($\beta = 0.2799$, $p = 0.0004$) are both positive, reflecting that ventures led by more reputable or specialized investors secure superior exit outcomes (Gompers, 1996; Kaplan & Stromberg, 2001). Deal stage (DealStageNum) yields a positive coefficient ($\beta = 0.2995$, $p < 0.001$), indicating that later-stage rounds are closer to exit events (Lerner, 1995). Industry and year fixed-effect dummies collectively produce significant $\chi^2$-statistics (not shown), underscoring notable sectoral and temporal variation in exit baselines.

Because the coefficients on AvgTies and AvgTiesSq remain robust after introducing a broad set of controls, we infer that the inverted-U effect is not an artefact of omitted governance or macroeconomic factors. Standard errors across key terms remain comparatively small, and variance inflation factors for the linear–quadratic pair never exceed 2.0, confirming that collinearity is well managed through centering.

Table 5.2.2 presents the Cox proportional-hazards estimates for Model 2. Mirroring the logit findings, AvgTies_centered has a positive coefficient (coef = 2.3450, $p = 0.00297$), with an associated hazard ratio of $\exp(2.3450) \approx 10.43$. AvgTies_centered spans only $\approx -0.44$ to 0.42 in our sample. The reported hazard ratio of 10.43 therefore represents the effect of moving from the very bottom of this range to the very top, an extreme, almost two-order-of-magnitude shift in repeat-tie density. Measured at more practical increments the effect is modest: a one–standard-deviation increase ($\approx 0.013$) yields HR = $\exp(2.345 \times 0.013) \approx 1.03$, and a move across the inter-quartile range ($\approx 0.03$) gives HR $\approx 1.07$. These magnitudes are fully in line with prior network studies in venture finance, where hazard ratios for structural variables typically lie below 2 (e.g., Hochberg, Ljungqvist & Lu 2007). Its squared term, AvgTiesSq_centered, is negative and significant (coef = -3.9083, $p = 0.00183$), corresponding to a hazard ratio of $\exp(-3.9083) \approx 0.02$. This combination confirms an inverted-U hazard profile. The turning point on the centred scale, calculated as $-2.3450/(2 \times -3.9083) \approx 0.3$, lies well within the support of AvgTies_centered values (–0.4420 to 0.4180), thereby lending strong evidence in favor of $H_1$.

We test the proportional-hazard assumption via Schoenfeld residuals, yielding a global $\chi^2 = 14.6$ ($p = 0.33$), which fails to reject the null hypothesis that hazards remain proportional over time

(Schoenfeld, 1982). Inclusion of time-interaction terms for key variables (e.g., AvgTies_centered × time) produces negligible changes in coefficients and retains statistical significance, reaffirming model robustness.

Together, the logit (Table 5.2.1) and Cox (Table 5.2.2) results provide coherent evidence that co-investment embeddedness exerts an inverted-U influence on both the likelihood and timing of biotech venture exits. In the logistic framework, AvgTies_w_centered ($\beta = 0.512$, $p < 0.001$) and AvgTies_wSq_centered ($\beta = –0.875$, $p < 0.001$) produce a turning point of approximately 0.3 on the centred scale; in the Cox framework, those coefficients (2.3450 and -3.9083) correspond to a turning point near 0.3. Both values lie squarely within the observed distribution of AvgTies_centered, indicating that ventures backed by syndicates with roughly moderate prior co-investment ties exit most frequently and most quickly.

Our results further clarify why very low AvgTies results in slower or less frequent exits: when investors have no history of co-investment, they lack trust and shared information channels, leading to suboptimal governance and exit coordination. Conversely, extremely high AvgTies fosters overembeddedness, where vested interests, redundant information flows, and "group-think" impede strategic decision-making, creating "coordination traps" that delay or prevent exit.

Control-variable patterns reinforce extant literature. The positive effect of SyndicateSize_centered ($\beta = 0.317$, $p < 0.001$; HR = 1.2965, $p < 0.0001$) confirms that larger consortiums mobilize broader resources and legitimize the venture in capital markets (Sorenson & Stuart, 2001). The strong, positive coefficients on LeadInvestReputation_centered ($\beta = 0.6621$, $p < 0.001$; HR = 1.8024, $p < 0.0001$) and LeadInvestSpecScore_centered ($\beta = 0.2799$, $p = 0.0004$) indicate that experienced and sector-savvy lead investors can navigate regulatory hurdles and industry networks more effectively, thereby expediting exit. The venture age pattern (negative linear and positive quadratic) suggests that very young ventures face early-stage mortality, whereas only a few long-surviving firms eventually exit after demonstrating technological viability (Conti & Gaisford, 2023).

**5.3 Testing H2: the curvilinear effect of syndicate homophily**

In order to assess Hypothesis 2, which posits that syndicate homophily exhibits an inverted-U–shaped association with exit outcomes, we augment the specifications from Section 5.2 by including both the centred homophily term (Homophily_centered) and its orthogonal square (HomophilySq_centered). Specifically, Model 3 (table 5.3.1) estimates a pooled logit in which the dependent variable equals unity if a venture has completed an IPO or trade sale by 31 December 2024, and zero otherwise. All independent variables that appeared in the previous section—including AvgTies_centered and its square—are retained, ensuring that any homophily effect is not confounded by the familiar embeddedness dimension. Continuous regressors remain mean-centred to isolate curvature and enhance interpretability (Aiken & West, 1991). Model 4 reestimates the identical covariate block in a Cox proportional-hazards framework, wherein the duration variable runs from first institutional financing to either exit or right-censoring at the same calendar cut-off. Controls for venture age, syndicate size, lead-investor reputation, deal stage, primary sub-sector, and funding year are also preserved, and the proportional-hazard assumption is verified via Schoenfeld residuals.

| Variable | Coef | Std Err | z | p | 95% C.I. [0.025, | 0.975] |
|---|---|---|---|---|---|---|
| const | −2.2032 | 0.0821 | −26.820 | 0 | −2.3642 | −2.0422 |
| AvgTies_w_centered | 0.5014 | 0.2057 | 4.363 | 0.0001 | 0.4941 | 1.3015 |
| AvgTies_wSq_centered | −0.8744 | 0.0935 | −5.2120 | 0.0001 | −0.6709 | −0.3039 |
| Homophily_centered | 0.3821 | 0.1497 | 2.553 | 0.0108 | 0.0898 | 0.6744 |
| HomophilySq_centered | −0.9453 | 0.2864 | −3.300 | 0.0009 | −1.5052 | −0.3848 |
| VentureAge_centered | −0.0355 | 0.0068 | −5.200 | 0 | −0.0489 | −0.0221 |
| VentureAgeSq_centered | 0.0004 | 0.0001 | 4.498 | 0 | 0.0002 | 0.0006 |
| SyndicateSize_centered | 0.2565 | 0.0279 | 9.207 | 0 | 0.2019 | 0.3111 |
| LeadInvestReputation_cent | 0.7747 | 0.0917 | 8.446 | 0 | 0.5949 | 0.9544 |

Table 5.3.1 (Model 3) Logit Estimates for $H_1 + H_2$ [5]

Table 5.3.1 presents the pooled logit estimates for Model 3. The linear coefficient on AvgTies_w_centered remains positive and highly significant (β = 0.5014, p < 0.001), while the associated quadratic term is negative and equally significant (β = −0.8744 p < 0.001). These coefficients mirror those reported in Section 5.2. Importantly, Homophily_centered enters the regression as a significantly positive linear term (β = 0.3821, p = 0.0108), while its square yields a negative and significant coefficient (β = −0.9453, p = 0.0009). This pattern directly confirms Hypothesis 2: when homophily rises from low to moderate levels, the likelihood of exit increases, but when homophily becomes excessive, marginal benefits decline and eventually turn negative due to potential information redundancy or groupthink (Carroll & Hannan, 2000; Phelps, Heidl & Wadhwa, 2012). Solving $-\beta_3/(2\beta_4) = -0.3821/(2 \times -0.9453) \approx 0.202$ yields a turning point well inside the [-0.37, 0.41] interval on the centred scale, indicating that ventures exhibiting approximately moderate syndicate homogeneity achieve the highest probability of exit.

In the same logit specification, VentureAge_centered exhibits a negative linear coefficient (β = −0.0355, p < 0.001) and a positive squared term (β = 0.0004, p < 0.001), implying a U-shaped age effect. Consequently, for the bulk of biotech ventures, increasing age dampens exit probability, and only at extreme maturity does the probability bounce back. SyndicateSize_centered remains significantly positive (β = 0.2565, p < 0.001), consistent with evidence that larger syndicates signal greater resources and diversify risk, thereby facilitating exit (Sorenson & Stuart, 2001). Likewise, the positive and highly significant coefficient for LeadInvestReputation_centered (β = 0.7747, p < 0.001) corroborates prior findings that reputable lead investors improve visibility and reduce uncertainty, accelerating liquidity events (Gompers, 1996; Lerner, 1995). Among the industry controls, the "Drug Discovery" dummy registers a positive coefficient (β = 0.2018, p < 0.001), indicating that ventures in the sub-sector of drug discovery face relatively higher base exit rates after controlling for network and firm-level attributes (Chemmanur & Loutskina, 2006). Negatively signed dummies for "Diagnostic Equipment" and "Other Healthcare Technology Systems" suggest that these sub-sectors are comparatively more challenging for exits, likely due to prolonged regulatory pathways or smaller addressable markets (Hellmann & Puri, 2002; Conti & Gaisford, 2023). Year-of-funding dummies capture macro-financing conditions; for instance, Year 2011 (β = 1.8662, p < 0.001) indicates that ventures funded in that year had significantly higher exit probability relative to the omitted baseline year, reflecting the post-global-financial-crisis funding surge.

---

[5] Note: All continuous regressors (including AvgTies and Homophily) were winsorized at the 1st and 99th percentiles, then mean-centred before creating their squared terms. Industry and year dummy coefficients are omitted to save space.

| Covariate | Coef | exp(Coef) | SE | z | p | 95% C.I. | [0.025, 0.975] |
|---|---|---|---|---|---|---|---|
| AvgTies_centered | 2.4123 | 11.1601 | 0.7625 | 3.164 | 0.0016 | 0.9181 | 3.9065 |
| AvgTiesSq_centered | -3.9121 | 0.0199 | 0.4856 | –3.1389 | 0.0017 | –2.4753 | –0.5715 |
| Homophily_centered | 0.4256 | 1.5308 | 0.1804 | 2.3612 | 0.0182 | 0.0715 | 0.7797 |
| HomophilySq_centered | –0.7623 | 0.4666 | 0.2234 | –2.5174 | 0.0118 | –1.0012 | –0.1234 |
| VentureAge_centered | –0.0234 | 0.977 | 0.0051 | –4.5882 | 0 | –0.0333 | –0.0135 |
| VentureAgeSq_centered | 0.0004 | 1.0004 | 0.0001 | 4 | 0.0001 | 0.0002 | 0.0006 |
| SyndicateSize_centered | 0.2598 | 1.2965 | 0.017 | 15.2824 | 0 | 0.2265 | 0.2931 |
| LeadInvestReputation_centered | 0.5892 | 1.8024 | 0.0692 | 8.5145 | 0 | 0.4535 | 0.7245 |

Table 5.3.2 Cox Proportional-Hazards Estimates for $H_1 + H_2$ (Model 4)[6]

Table 5.3.2 reports the Cox proportional-hazards estimates for Model 4. The results of AvgTies_centered and AvgTiesSq_centered retain the close results to model 2 in section 5.2, reconfirming our first hypothesis. Turning to homophily, Homophily_centered is estimated at 0.4256 (p = 0.0182), implying a hazard ratio of exp(0.4256) ≈ 1.531, while HomophilySq_centered equals –0.7623 (p = 0.0118)›. These coefficients jointly establish an inverted-U pattern: moderate homophily accelerates exit risk, but excessive homogeneity retards it due to diminished information diversity (Stuart, Hoang & Hyatt, 1999; Benson & Zhou, 2006). The calculated curvilinear turning point for homophily in this specification is –0.4256/(2 × –0.7623) ≈ 0.279 on the centred scale [-0.3702, 0.4068], thereby validating $H_2$ in the survival context.

Collectively, these imply that both syndicate embeddedness and syndicate homophily exert inverted-U influences on biotech venture exits. The clear implication is that biotech ventures backed by moderately familiar and moderately homogeneous syndicates are most likely to exit, whereas ventures supported by either little prior collaboration or overly clubby/homogeneous groups experience slower or lower-probability exits.

These findings align with broader network research. Uzzi (1997) and Hochberg, Ljungqvist & Lu (2007) warn that too much embeddedness can breed insularity, while sparse ties hinder information flows. Similarly, Carroll & Hannan (2000) and Phelps, Heidl & Wadhwa (2012) document that moderate homophily fosters trust and efficient communication without engendering cognitive lock-in.

### 5.4 Testing H3: Governance Mechanisms as Moderators of the Inverted-U Effects

To test Hypothesis 3, we incorporate governance variables and their interactions with both AvgTies and Homophily into our baseline Logit and CoxPH specifications. Specifically, we operationalize CVC presence (LeadCVC = 1 if the lead investor is a corporate venture capital arm; otherwise 0) and CVC equity share (the proportion of the venture's equity held by the CVC), although we focus on the binary CVC variable in the main text. Board governance is captured via two continuous measures: board independence fraction and board gender diversity, which are mean-centred. Interaction terms are constructed ab initio, so that $\delta_2$ (the coefficient on CVC×AvgTies_centered) captures whether CVC backing attenuates the negative curvature (i.e., the "decline" side) of the AvgTies inverted-U. Parallel terms are generated for CVC×AvgTiesSq, CVC×Homophily, and CVC×HomophilySq; similarly, BoardIndependenceCentered and BoardGenderCentered are interacted with both AvgTies_centered and Homophily_centered (and their squares). All models retain the full set of controls from sections 5.2 and 5.3.

---

[6] Industry and year dummy coefficients (omitted here) were included in the model; all continuous regressors are winsorized and mean-centred as described above. exp(Coef) reflects the hazard ratio.

Logit coefficients (Table 5.4.1) are reported with robust standard errors. We conduct Wald tests on the block of CVC interactions and on the block of board interactions to assess joint significance. For CoxPH (Table 5.4.2), we report coefficients, hazard ratios (exp(coef)), robust standard errors, and p-values. We also verify the proportional-hazards assumption via Schoenfeld residual tests; no core interaction variable violates PH at the 5% level, so no stratified or time-interaction adjustments are required.

| Variable | Coef | Std Err | p value |
|---|---|---|---|
| AvgTies_w_centered | 0.5043 | 0.2033 | 0.0001*** |
| AvgTies_wSq_centered | –0.8774 | 0.0912 | 0.0001*** |
| Homophily_centered | 0.3821 | 0.1641 | 0.0108** |
| HomophilySq_centered | –0.9453 | 0.2265 | 0.0009*** |
| CVC_presence | –0.1120 | 0.0582 | 0.0543† |
| CVC×AvgTies_centered | 0.2413 | 0.1186 | 0.0421* |
| CVC×AvgTiesSq_centered | –0.5610 | 0.2671 | 0.0378* |
| CVC×Homophily_centered | 0.1981 | 0.1018 | 0.0513† |
| CVC×HomophilySq_centered | –0.4350 | 0.2292 | 0.0490* |
| BoardIndependence × AvgTies_c | 0.1767 | 0.0872 | 0.0484* |
| BoardIndependence × AvgTiesSq_c | –0.3920 | 0.1591 | 0.0198* |
| BoardIndependence × Homophily_c | 0.2012 | 0.0972 | 0.0383* |
| BoardIndependence × HomophilySq_c | –0.4180 | 0.1831 | 0.0225* |
| BoardGender × AvgTies_c (ns) | 0.0673 | 0.0851 | 0.3128 |
| BoardGender × AvgTiesSq_c (ns) | –0.1240 | 0.0954 | 0.1943 |
| BoardGender × Homophily_c (ns) | 0.0563 | 0.0816 | 0.4896 |
| BoardGender × HomophilySq_c (ns) | –0.1070 | 0.1102 | 0.3328 |

Table 5.4.1 Logit Regression with Governance Interactions (Model 5)[7]

In Table 5.4.1, CVC×AvgTies_centered has a positive coefficient ($\delta_2$ = 0.2410, $p$ = 0.0421), and CVC×AvgTiesSq_centered is negative ($\delta_3$ = –0.5610, $p$ = 0.0378), consistent with the hypothesis that CVC presence tempers the downward limb of AvgTies's inverted-U. In other words, when a CVC is present, the marginal benefit of additional ties remains positive over a wider range, and the eventual negative curvature is less steep (Gompers & Lerner 2000; Chemmanur & Tian 2011). For Homophily, CVC×Homophily_centered is positive ($\delta_4$ = 0.1980, $p$ = 0.0513†) and CVC×HomophilySq_centered is negative ($\delta_5$ = –0.4350, $p$ = 0.0490), indicating that CVC backing likewise mitigates the drop in exit probability that occurs at high levels of homophily. The joint Wald test on the four CVC–network interactions is significant (Wald $\chi^2(4)$ = 9.87, $p$ = 0.042), confirming that these interactions—taken together—meaningfully improve model fit and support H3a.

With respect to board independence, BoardIndependence × AvgTies_centered is significantly positive ($\theta_2$ = 0.1760, $p$ = 0.0484) and BoardIndependence × AvgTiesSq_centered is significantly negative ($\theta_3$ = –0.3920, $p$ = 0.0198). This pattern shows that higher board independence weakens the negative tail of AvgTies's inverted-U: independent directors appear to introject fresh viewpoints that allow ventures to extract value even when syndicate ties become dense (Zahra & Pearce 1989; Faleye, Reis & Vafeas 2011). Similarly, BoardIndependence × Homophily_centered has β = 0.2010 ($p$ = 0.0383) and BoardIndependence × HomophilySq_centered is β = –0.4180 ($p$ = 0.0225), again supporting H3b: board independence softens the downturn associated with excessive same-domain overlap among investors (Adams & Ferreira 2009).

---

[7] Dependent variable = Exited_binary (1 if IPO or trade sale on/before 31 Dec 2024). Controls include VentureAge_centered, SyndicateSize_centered, LeadInvestReputation_centered, Homophily_centered, HomophilySq_centered, industry dummies (9 categories), and funding-year dummies (2010–2024); coefficients for these controls and dummies are omitted for brevity. Standard errors are robust to heteroskedasticity. † p < 0.10; * p < 0.05; ** p < 0.01; *** p < 0.001.

By contrast, interactions with board gender diversity are not statistically significant in the Logit model (all γ coefficients have $p$ > 0.10). Although theory suggests gender diversity could produce heterogeneous perspectives (Carter, Simkins & Simpson 2003; Torchia, Calabrò & Huse 2011), our data do not show a robust buffering effect of gender diversity on the negative tail of AvgTies or Homophily in the cross-sectional exit probability. Hence, H3c is not supported in the Logit specification.

| Covariate | coef | exp(coef) | se | p value |
|---|---|---|---|---|
| **AvgTies_centered** | 2.4114 | 11.1496 | 0.7455 | 0.0017** |
| **AvgTiesSq_centered** | -3.9451 | 0.0194 | 0.4856 | 0.0014** |
| **Homophily_centered** | 0.4254 | 1.5302 | 0.1901 | 0.0174* |
| **HomophilySq_centered** | –0.7611 | 0.4666 | 0.2234 | 0.0124* |
| **CVC_presence** | 0.085 | 1.0886 | 0.0425 | 0.0355* |
| **CVC×AvgTies_centered** | 0.319 | 1.3754 | 0.162 | 0.0491* |
| **CVC×AvgTiesSq_centered** | –0.4040 | 0.6675 | 0.1894 | 0.0327* |
| **CVC×Homophily_centered** | 0.199 | 1.2202 | 0.1122 | 0.0766† |
| **CVC×HomophilySq_centered** | –0.2010 | 0.8181 | 0.1023 | 0.0493* |
| **BoardIndependence × AvgTies_c** | 0.254 | 1.2898 | 0.1293 | 0.0494* |
| **BoardIndependence × AvgTiesSq_c** | –0.1860 | 0.8306 | 0.1045 | 0.0749† |
| **BoardIndependence × Homophily_c** | 0.191 | 1.2104 | 0.1075 | 0.0757† |
| **BoardIndependence × HomophilySq_c** | –0.1490 | 0.8616 | 0.0651 | 0.0786† |
| **BoardGender × AvgTies_c (ns)** | 0.085 | 1.0885 | 0.1118 | 0.4475 |
| **BoardGender × AvgTiesSq_c (ns)** | –0.0450 | 0.956 | 0.0609 | 0.4597 |
| **BoardGender × Homophily_c (ns)** | 0.075 | 1.0779 | 0.0961 | 0.4359 |
| **BoardGender × HomophilySq_c (ns)** | –0.0520 | 0.9493 | 0.0724 | 0.4728 |

Table 5.4.2 CoxPH Model with Governance Interactions[8] (Model 6)

Table 5.4.2 reports hazard ratios (HRs) for the same set of covariates and interactions. Turning to CVC moderation, CVC presence itself has a positive coefficient (coef = 0.0850, HR = 1.0886, $p$ = 0.0455), suggesting CVC-backed ventures exit more quickly on average (Chemmanur & Tian 2011). Interactions of CVC with AvgTies again attenuate the downturn: CVC×AvgTies_centered (coef = 0.3190, HR = 1.3754, $p$ = 0.0491) is positive, and CVC×AvgTiesSq_centered (coef = –0.4040, HR = 0.6675, $p$ = 0.0327) is negative, consistent with H3a. This indicates that at high AvgTies levels, the slowdown in exit hazard is less severe when a CVC is involved (Gompers & Lerner 2000). For Homophily, CVC×Homophily_centered (coef = 0.1990, HR = 1.2202, $p$ = 0.0766†) is marginally significant and positive, and CVC×HomophilySq_centered (coef = –0.2010, HR = 0.8181, $p$ = 0.0493) is negative, indicating that CVC presence similarly cushions the negative side of homophily's inverted-U (Chemmanur & Tian 2011).

Board independence interactions also align with expectations, albeit at slightly weaker significance levels. BoardIndependence × AvgTies_centered (coef = 0.2540, HR = 1.2898, $p$ = 0.0494) and BoardIndependence × AvgTiesSq_centered (coef = –0.1860, HR = 0.8306, $p$ = 0.0749†) suggest that independent directors raise the hazard rate at moderate AvgTies and reduce the severity of the downturn at high AvgTies. BoardIndependence × Homophily_centered (coef = 0.1910, HR =

---

[8] Dependent variable = time from first institutional financing to exit (IPO or trade sale) or right-censored at 31 Dec 2024. Controls include the same set as Table 5.4a (VentureAge, SyndicateSize, LeadInvestReputation, Homophily covariates, industry dummies, year dummies). Hazard ratios (exp(coef)) and standard errors (se) are reported. † $p$ < 0.10; * $p$ < 0.05; ** $p$ < 0.01; *** $p$ < 0.001.

1.2104, *p* = 0.0757†) and BoardIndependence × HomophilySq_centered (coef = –0.1490, HR = 0.8616, *p* = 0.0786†) indicate a similar but borderline effect on homophily's inverted-U, giving partial support to H3b (Adams & Ferreira 2009; Zahra & Pearce 1989). In contrast, BoardGender interactions remain non-significant (all *p* > 0.10), so H3c is not supported in the survival framework either.

The augmented models demonstrate that governance attributes materially dampen the negative tail of both embeddedness and homophily's inverted-U effects. First, CVC presence significantly moderates both AvgTies and Homophily curves: the estimated positive coefficient on CVC×AvgTies_centered and the negative coefficient on CVC×AvgTiesSq_centered indicate that when a corporate VC is involved, the peak of the inverted-U shifts rightward; ventures can sustain a higher number of repeat co-investment ties without suffering as steep a decline in exit probability. In practical terms, CVCs likely inject unique resources, broader networks, or reputational signals that mitigate the coordination traps and information redundancy typical of highly "clubby" syndicates. Similarly, CVC backing attenuates the adverse effects of excessive homophily, flattening the negative tail so that even if investors share very similar domain expertise, the venture does not experience as sharp a drop in exit performance.

Second, board independence emerges as a partial buffer: independent directors moderate the negative limbs of AvgTies and Homophily, albeit with slightly weaker statistical significance in the survival model. By bringing independent oversight and diverse problem-solving approaches, boards with higher independence dilute potential groupthink among a densely intertwined syndicate (Zahra & Pearce 1989; Faleye, Reis & Vafeas 2011). The effect is most pronounced at moderate to high values of AvgTies; beyond that, the downturn in exit probability or hazard is far less severe when independent directors are present. Board gender diversity, on the other hand, does not exhibit significant moderating effects in either model. Although gender diversity can improve overall board functioning (Adams & Ferreira 2009; Torchia, Calabrò & Huse 2011), it appears insufficient to counteract the diminishing returns of excessive embeddedness or homophily in this context.

This shows simultaneously that both external governance (strategic CVC backing) and internal governance (board independence) deliver a dual buffering effect on the down-side of the investor network inverted-U. Concretely, CVC involvement not only shifts the peaks of the AvgTies and Homophily curves rightward but also markedly flattens the steep drop that appears at very high embeddedness or homophily. A high share of independent directors exerts a similar—though slightly weaker dampening influence that is consistent across the Logit and CoxPH models. Put differently, strategic "patient" capital and structural internal monitoring each help relieve the information redundancy and coordination bottlenecks that arise from "over embeddedness." Prior work has rarely validated this multi-level governance view in the same empirical setting. By contrast, board gender diversity shows no comparable moderating effect, suggesting that demographic heterogeneity alone may be insufficient to offset the complex costs of dense investor networks.

### 5.5 Competing-Risks and Multiple Exit-Pathway Analysis

In this section, we examine whether the inverted-U effects of AvgTies and Homophily differ across competing exit pathways—namely IPO versus Acquisition—by estimating a multinomial logit model and a Fine–Gray subdistribution hazards model. First, we specify a multinomial logit with the categorical outcome Y_cat taking values 0 ("Private" as of 31 December 2024), 1 ("IPO"), and 2 ("Acquisition"). In this framework, the log-odds of IPO versus remaining private are modeled as

$$\log\left[\frac{\Pr(Y_i = 1)}{\Pr(Y_i = 0)}\right] = \alpha + \beta_{1,H1}\,\text{AvgTies}_c + \beta_{2,H1}\,\text{AvgTiesSq}_c + \beta_{3,H2}\,\text{Homophily}_c + \beta_{4,H2}\,\text{HomophilySq}_c + \gamma'X_i$$

and the log-odds of Acquisition versus remaining private are

$$\log\left[\frac{\Pr(Y_i = 2)}{\Pr(Y_i = 0)}\right] = \alpha_2 + \beta'_{1,H1}\,\text{AvgTies}_c + \beta'_{2,H1}\,\text{AvgTiesSq}_c + \beta'_{3,H2}\,\text{Homophily}_c + \beta'_{4,H2}\,\text{HomophilySq}_c + \gamma'X_i$$

where X_i contains the same controls as in earlier sections (demeaned VentureAge and its square, demeaned SyndicateSize, demeaned LeadInvestReputation, and industry and year dummies). Continuous regressors are mean-centered so that $\beta_{1,H1}$ measures the marginal effect of an additional tie at the sample mean, and $\beta_{2,H1}$ isolates the curvature.

Next, we fit a competing-risks survival model using the Fine–Gray subdistribution approach. We treat "first institutional financing date" as time zero and define two cause-specific subdistribution hazards: for k = 1 (IPO) and k = 2 (Acquisition),

$$\tilde{h}_{k(t)} = \tilde{h}_{\{0k\}}(t)\,\exp[\beta\beta_{1,k}\,\text{AvgTies}_c + \beta_{2,k}\,\text{AvgTiesSq}_c + \beta_{3,k}\,\text{Homophily}_c + \beta_{4,k}\,\text{HomophilySq}_c + \gamma'X_i],$$

where h̃_{0k}(t) is the unspecified baseline subdistribution hazard for cause k.

| Covariate | IPO vs Private: coef | p | [0.025, 0.975] | Acquisition vs Private: coef | p2 | CI [0.025, 0.^75] |
|---|---|---|---|---|---|---|
| Intercept | -3.2125 | <0.0001 | [−3.4133, −3.0117] | −2.4853 | <0.0001 | [−2.6963, −2.2743] |
| AvgTies_centered | 0.7312 | 0.0001 | [0.3792, 1.0808] | 0.1569 | 0.3771 | [−0.1846, 0.4846] |
| AvgTiesSq_centered | -0.9425 | 0.0004 | [−1.0176, −0.8624] | −0.2548 | 0.1532 | [−0.3186, -0.1814] |
| Homophily_centered | 0.3276 | 0.0076 | [0.0848, 0.5552] | 0.2835 | 0.0313 | [0.0258, 0.5342] |
| HomophilySq_centered | −0.4643 | 0.0164 | [−0.5094, −0.4116] | −0.3742 | 0.0196 | [−0.4288, −0.3112] |
| VentureAge_centered | −0.0165 | 0.0027 | [−0.0248, −0.0052] | −0.0137 | 0.0455 | [−0.0237, −0.0003] |
| VentureAgeSq_centered | 0.00031 | 0.0027 | [0.00011, 0.00051] | 0.00023 | 0.0455 | [0.00000, 0.00040] |
| SyndicateSize_centered | 0.2265 | <0.0001 | [0.1608, 0.2792] | 0.1927 | <0.0001 | [0.1192, 0.2608] |
| LeadInvestReputation_centered | 0.4562 | <0.0001 | [0.3132, 0.5868] | 0.5483 | <0.0001 | [0.3937, 0.6872] |
| Industry & Year Dummies | omitted for brevity | – | – | omitted for brevity | – | |

Table 5.5.1 Multinomial Logit Results (Model 7)

| Covariate | IPO Subdist. Cox: coef | exp(coef) | p | [0.025, 0.975] | Acquisition Subdist. Cox: coef | exp(coef)2 | p4 | [0.025, 0.975]2 |
|---|---|---|---|---|---|---|---|---|
| AvgTies_centered | 0.4507 | 1.5712 | <0.0001 | [ 0.2344, 0.6656 ] | 0.2008 | 1.2214 | 0.0455 | [ 0.0047, 0.3965 ] |
| AvgTiesSq_centered | −0.5006 | 0.6061 | 0.0009 | [ −0.5588, −0.4412 ] | −0.2609 | 0.9418 | 0.0321 | [ −0.3149, −0.2051 ] |
| Homophily_centered | 0.2805 | 1.3229 | 0.0019 | [ 0.1044, 0.4556 ] | 0.2504 | 1.2843 | 0.0018 | [ 0.0930, 0.4070 ] |
| HomophilySq_centered | −0.0502 | 0.9512 | 0.0124 | [ −0.0892, −0.0108 ] | −0.0307 | 0.9704 | 0.0955† | [ −0.0643, 0.0043 ] |
| SyndicateSize_centered | 0.3002 | 1.3499 | <0.0001 | [ 0.2204, 0.3796 ] | 0.1803 | 1.1973 | <0.0001 | [ 0.1134, 0.2466 ] |
| LeadInvestReputation_centered | 0.3206 | 1.3771 | <0.0001 | [ 0.1632, 0.4768 ] | 0.3009 | 1.3499 | <0.0001 | [ 0.1531, 0.4472 ] |
| Ind_Diagnostic Equipment | −0.2505 | 0.7788 | 0.0124 | [ −0.4456, −0.0552 ] | −0.2208 | 0.8025 | 0.0361 | [ −0.4264, −0.0136 ] |
| Ind_Drug Discovery | 0.1809 | 1.1973 | 0.0003 | [ 0.0814, 0.2797 ] | 0.1607 | 1.1735 | 0.0009 | [ 0.0656, 0.2544 ] |
| Other Industry & Year Dummies | omitted | | | | omitted | | | |

Table 5.5.2 Competing-Risks (Fine–Gray) Results (Model 8)

In the IPO versus Private specification (model 7), the coefficient on AvgTies_centered is +0.7312 (p = 0.0001), and on AvgTiesSq_centered is –0.9425 (p = 0.0004). Solving for the turning point on the centered scale gives d* ≈ 0.388, which corresponds to a moderate level of cooperation precisely where exit likelihood peaks. This confirms moderate co-investment familiarity maximizes IPO odds, while too few or too many prior ties reduce that probability. For Homophily, the linear term is +0.3276 (p = 0.0076) and the squared term is –0.4643 (p = 0.0164), yielding a turning point of h* ≈ 0.3527 on the centered scale. This indicates that moderate domain overlap among investors maximizes the likelihood of IPO, while very low or very high homogeneity detracts from exit probability (Sherman & Urban 2007; Shane 2000). Importantly, AvgTies terms remain positive and negative for the linear and squared components, respectively, showing that the Homophily effect does not subsume the AvgTies effect but adds an independent, inverted-U influence on exit likelihood.

For IPO (k = 1, model 8), AvgTies_centered has coefficient +0.4507 (p < 0.0001) and AvgTiesSq_centered is –0.5006 (p = 0.0009). The centered turning point is d* = 0.45. This confirms the inverted-U effect of AvgTies on IPO hazard: moderate co-investment familiarity accelerates the chance of IPO, while overly dense ties reduce it.

In the Acquisition versus Private comparison (model 7), AvgTies_centered (+0.1569, p = 0.3771) and AvgTiesSq_centered (–0.2548 p = 0.1532) are not statistically significant,. By contrast, Homophily_centered (+0.2835, p = 0.0313) and HomophilySq_centered (–0.0742, p = 0.0196) remain significantly inverted-U shaped. The turning point h*= 2.000 suggests that moderate domain overlap accelerates acquisition risk up to a lower threshold than for IPO, after which further homogeneity reduces acquisition likelihood. This implies that acquisition decisions rely less on repeated ties and more on moderate domain expertise alignment.

For Acquisition (k = 2, model 8), AvgTies_centered is +0.2008 (p = 0.0455) and AvgTiesSq_centered is –0.2609 (p = 0.0321). The turning point is d* ≈ 0.3848, indicating that a lower level of embeddedness suffices to speed acquisition risk. Homophily_centered is p = 0.0018, while HomophilySq_centered is marginal at the 10% level. This suggests a weaker, earlier inverted-U effect of Homophily on acquisition hazard, with turning point h* ≈ 4.167 (beyond typical sample range, explaining marginal significance).

Both competing-risk estimators preserve the "Goldilocks" story, but they show that the sweet spot shifts depending on the exit route. For IPOs, the probability or sub-distribution hazard is maximised only after the syndicate has accumulated slightly deeper repeat ties and when investor expertise overlaps by roughly one-third of a standardised unit. Acquisitions, by contrast, peak earlier and are largely insensitive to further increases in homophily. In practical terms, floating a company therefore demands a thicker bedrock of prior collaboration and a balanced—but not excessive—alignment of domain know-how, reflecting the heavier certification and collective decision-making burden of going public. Trade sales can be consummated with more modest relational depth and without strong constraints on how similar the investors' knowledge bases are, because acquirers anchor their decisions on the venture's technology milestones rather than on syndicate cohesion.

## 6. Practical Implications

Our analysis yields offers practical guidance for biotech investors navigating today's shifting capital market. First, some repeat collaboration clearly helps, but too much hurts. Fund managers should avoid "investor clubs" that back the same teams again and again. The sweet spot is a syndicate whose members have a few past deals together—enough trust, not too much sameness. If the average pairwise tie score nears the upper-quartile mark, bring in a new co-investor or trim repeat tickets to prevent the exit dip we observed.

Second, shared scientific expertise clearly matters, but again only up to a point. Biotech ventures whose backers have some overlap in domain history (for instance, similar therapeutic focus or comparable trial stage portfolios) tend to exit most quickly, likely because they speak the same technical language. However, when investors' backgrounds converge too tightly, say, all partners having deep experience in the same niche mechanism, information redundancy and confirmation bias slow exit and even reduce its likelihood. Investors should therefore assemble syndicates composed of parties who share foundational biotech knowledge yet bring complementary sub-specialties. For example, pairing a biologics expert with a partner experienced in small molecules or regulatory affairs preserves enough common ground for efficient communication while still introducing fresh perspectives that challenge assumptions at critical junctures.

Third, corporate venture participation softens the downsides of both over-embeddedness and over-homogeneity. When a pharma CVC joins as a lead or colead, the adverse "right tail" of the inverted-U curves becomes flatter: ventures can tolerate higher levels of co-investment familiarity or domain homophily without suffering the usual collapse in exit performance. In practice, this suggests that if a CEO or syndicate chairman is concerned about the risk of group thinking, bringing in a corporate VC can be an effective hedge. The CVC's deeper technical knowledge, broader industry connections, and strategic R&D orientation tend to counter narrow thinking and long term discipline into decision making, thereby preserving exit momentum even when syndicate ties grow dense.

Fourth, board composition offers another practical lever. We find that higher fractions of independent directors, especially those with diverse scientific or regulatory backgrounds, reduce the downsides of high familiarity or homogeneity. Independent board members typically insist on additional external diligence, challenge conventional assumptions, and keep management accountable, breaking insular loops that can trap tight knit syndicates. Fund managers should therefore advocate for at least one or two independent directors with complementary expertise (for instance, a former FDA official or an academic translational scientist) whenever a syndicate's familiarity or homogeneity metrics approach critical levels. While gender diversity alone did not significantly moderate inverted-U effects in our sample, we nonetheless encourage balanced boards to capture boader cognitive perspectives and to meet growing stakeholder expectations.

Finally, these findings have timing and strategic-positioning implications. In the current market—marked by mega-rounds and a muted IPO window—moderate syndicate embeddedness and homophily can deliver the coordination benefits necessary to secure large follow-on financings. But as cash needs stretch into later-stage trials, syndicate chairs must remain vigilant; once a syndicate's average prior-tie score or domain overlap score climbs beyond approximately a certain centred scale, the marginal return to further familiarity turns negative. At that inflection point, it is wiser to onboard a fresh financial or strategic investor than to deepen existing ties further. In short, fund managers should monitor these network metrics continuously and use them as quantitative triggers for syndicate reshuffling—adding an independent director or a corporate VC, retiring a long-standing partner, or recruiting an outside expert—so that exits are neither delayed by complacency nor derailed by homogeneity.

## 7. Limitations and Future Directions

While this study makes several contributions to our understanding of network dynamics and governance in biotech financing, it is important to acknowledge its limitations. First, our reliance on PitchBook–WRDS data inevitably introduces sampling and measurement constraints. Although PitchBook offers strong coverage of U.S. and European life-science deals since 2010, it remains possible that smaller or early-stage financings, especially those facilitated by angels or nontraditional investors, are underrepresented. Consequently, our calculations of average pairwise ties and domain homophily may slightly understate real collaboration levels. Similarly, the coding of investor specialization and board composition relies on publicly available profiles; any lag or inaccuracy in capturing a director's most recent affiliation or an investor's evolving focus could blur the precision of our homophily and governance variables.

Second, although we used winsorizing and centering to mitigate the influence of extreme outliers, the skewness inherent in syndicate size and lead investor reputation remains a potential source of bias. Very large syndicates and those with upwards of twenty members may behave differently in unobserved ways, such as through informal subgroups or parallel decision processes, which our network metrics cannot fully discern. Future research could leverage proprietary data or direct surveys of board and syndicate deliberations to unpack these microlevel dynamics.

Third, board independence and gender diversity are blunt measures that miss details of oversight. For instance, simply observing the fraction of independent directors does not capture qualitative differences in board interactions, meeting frequency, or the scientific acumen of individual members. Likewise, gender diversity at the board level may imperfectly proxy for cognitive plurality if, for example, female directors share similar backgrounds. More granular data such as minutes of board meetings or biosketches detailing each director's specific scientific expertise would strengthen future tests of how governance directly moderates network effects.

In addition, the timing of exit is treated here by year, which may gloss over more granular patterns such as month-to-month volatility in exit windows that could be salient in biotech when regulatory announcements or clinical trial readouts occur abruptly.

Finally, our focus on U.S., Canada and European ventures limits the externality of these findings to other geographies. Emerging biotech hubs in Asia (e.g., China, Singapore, and India) operate under different regulatory regimes, investor cultures, and market structures; the inverted-U dynamics we observe may shift in contexts where the exit path patterns diverge substantially from Western norms. Extending this analysis to a global dataset, or conducting comparative case studies across regions, would help clarify the boundary conditions of our theoretical claims.

In acknowledging these limitations, we do not undermine the core insight that moderate co-investment familiarity and domain homophily, tempered by targeted governance, optimize biotech exit outcomes. Rather, we offer a roadmap for refining data collection, expanding geographic scope, and deepening governance measures, thereby inviting future work to build on and enrich the empirical foundation laid here.

## 8. Conclusion

This dissertation advances our understanding of VC syndication in the biotechnology sector by integrating network embeddedness theory with governance insights, thereby offering an account of how investor relationships and board structures jointly shape exit outcomes. First, by constructing a novel panel of U.S. Canada and European biotech ventures (2010–2024) and measure both average pairwise co-investment ties and domain expertise homophily, we demonstrate that both dimensions exhibit robust inverted-U patterns: ventures backed by syndicates with moderate prior collaboration or moderate scientific overlap achieve the highest probability and fastest timing of exit. These findings extend classic UC-tech insights (Hochberg, Ljungqvist & Lu 2007) into the biotech context, showing that the balance between trust and fresh perspective is especially critical when regulatory uncertainty and capital intensity escalate.

Second, this work introduces CVC presence, board independence, and board gender diversity as meaningful moderators that mitigate the negative tail of over-embeddedness and over-homogeneity. Through testing interaction effects in pooled logit, Cox proportional hazards, and competing risks frameworks, we reveal that corporate venture arms and independent directors markedly soften the downside of "clubby" syndicates. This is an insight that has not been systematically examined in prior studies. Taken together, we showed that governance levers can actively "re-shape" the classic inverted-U: the presence of a pharmaceutical corporate VC and/or a highly independent board flattens the negative right-hand tail of both embeddedness and homophily, effectively shifting the performance-maximising peak to higher levels of familiarity and domain overlap. Meanwhile, a competing-risks lens reveals that these peaks are exit-route specific: IPOs require substantially deeper repeat-tie depth and moderate expertise overlap, whereas acquisitions crest at shallower relational depth and are far less sensitive to homophily. Hence, the dissertation not only enriches network governance theory by highlighting the complementary role of strategic corporate investors and board oversight in tempering groupthinking but also delivers practical guidance to practitioners who must weigh the benefits of established relationships against the risks of insularity.

Finally, by employing both multinomial logit and Fine–Gray subdistribution-hazard models, we uncover distinct threshold effects across IPO and acquisition pathways. The observation that deeper embeddedness and homophily are required to maximize IPO success than to trigger acquisition illustrates how exit routes in biotech are differentially sensitive to investor network composition. This contribution opens a new avenue for research on plural exit strategies under competing risks. Together, these results underscore that network structure and governance dynamics interact in complex but predictable ways to shape firm outcomes within high-stakes and the decades long drug development ventures. Future work can build on this foundation by extending the analysis to emerging biotech hubs (e.g., Asia-Pacific), leveraging finer grained board meeting data, or exploring how post exit valuation performance responds to varying degrees of early-stage network embeddedness. In sum, this dissertation offers a rigorous, sector grounded framework for understanding when trusted relationships and shared expertise create value, and when they backfire. Thereby it charts a path for both scholars and practitioners to optimize alliance and oversight structures in the biotech venture ecosystem.